\documentclass[sigconf]{acmart}

\usepackage{lipsum}   
\usepackage{comment}
\usepackage{longtable}
\usepackage{multirow}
\usepackage{tikz}
\usepackage{amsthm}
\usepackage{algorithmic}
\usepackage{textcomp}
\usepackage{filecontents}
\usepackage{algorithm}
\usepackage{listings}
\usepackage{xurl}

\theoremstyle{definition}

\definecolor{codegreen}{rgb}{0,0.6,0}
\definecolor{codegray}{rgb}{0.5,0.5,0.5}
\definecolor{codepurple}{rgb}{0.58,0,0.82}
\definecolor{backcolour}{rgb}{0.95,0.95,0.92}

\lstdefinestyle{mystyle}{
  backgroundcolor=\color{backcolour}, 
  commentstyle=\color{codegreen},
  keywordstyle=\color{magenta},
  numberstyle=\tiny\color{codegray},
  stringstyle=\color{codepurple},
  basicstyle=\ttfamily\footnotesize,
  breakatwhitespace=false,         
  breaklines=true,                 
  captionpos=b,                    
  keepspaces=true,                 
  numbers=left,                    
  numbersep=5pt,                  
  showspaces=false,                
  showstringspaces=false,
  showtabs=false,                  
  tabsize=2
}

\copyrightyear{2026}
\acmYear{2026}
\setcopyright{cc}
\setcctype{by}
\acmConference[AISec '26]{19th Workshop on Artificial Intelligence and Security}{November 15--19, 2026}{The Hague, Netherlands}
\acmBooktitle{19th Workshop on Artificial Intelligence and Security (AISec '26), November 15--19, 2026, The Hague, Netherlands}
\acmDOI{10.1145/3847352.3848102}
\acmISBN{979-8-4007-3031-3/2026/11}

\begin{document}

\title{Semi-Automated Detection of Gaps in LLM Security Knowledge}
\author{Shufan Chai}
\authornote{The first 2 authors contributed equally. Most of this work was done while students at Northeastern University in Oakland, CA, USA.}
\affiliation{%
  \institution{Northeastern University}
  \city{Oakland}
 \state{CA}
 \country{USA}
}
\email{chai.shu@northeastern.edu}

\author{Liangliang Sun}
\authornotemark[1]
\affiliation{%
  \institution{Northeastern University}
 \city{Oakland}
  \state{CA}
  \country{USA}
}
\email{sun.liang@northeastern.edu}

\author{Jessica Staddon}
\affiliation{%
  \institution{Northeastern University}
  \city{Oakland}
 \state{CA}
  \country{USA}
}
\email{j.staddon@northeastern.edu}

\begin{abstract}
Large language models (LLMs) are increasingly used for a range of software, hardware and human-centered security tasks. 
Consequently, LLM performance on security tasks is an active area of measurement and research, often with a focus on identifying areas in which LLM security ``knowledge'' may be insufficient. Popular strategies for identifying LLM security knowledge gaps include building corpora of challenge questions or task benchmarks, strategies that require substantial manual work and security expertise to design and execute. We introduce a partially-automated method for assessing LLM knowledge of a security area. The method uses authoritative information from Consumer Protection Agencies (CPAs) to identify instability in LLM responses that can be indicative of knowledge gaps. We demonstrate the method for 2 security topics, identity theft and impostor scams, and 5 LLMs in 2 leading LLM families, Gemini and GPT, using publicly available information about identity theft and impostor scams from 6 CPAs.
The method distinguishes between models that have and don't have sufficient knowledge to accurately identify the security topics in text narratives. Two models, GPT 5.5 and Gemini 3.1 Pro, demonstrate solid knowledge of impostor scams, whereas all models appear to have gaps in knowledge of identity theft.
\end{abstract}



\maketitle

\section{Introduction}
\label{sec:intro}

Large language models (LLMs) are increasingly being used for a variety of security\footnote{In this paper we use ``security'' in the same broad sense as the FTC and include issues of consumer safety and data privacy (\url{https://www.ftc.gov/about-ftc/bureaus-offices/bureau-consumer-protection/about-bureau-consumer-protection}).} tasks \cite{xu2024large}. For example, LLMs are used to generate security assertions for hardware \cite{kande2024security}, detect software vulnerabilities \cite{zhang2023well} and generate patches \cite{kulsum2024case}, protect end users from scams \cite{shen2025warned} and advise users on a variety of security and privacy topics \cite{chen2023can}. These use cases may involve direct user interaction with LLMs or the development of LLM-enabled agents to complete security tasks on behalf of users \cite{shao2026towards}.

The breadth of use cases suggests LLMs may be able to democratize access to specialized security knowledge across users and organizations, provided LLMs have solid security ``knowledge''\footnote{For simplicity of exposition, we characterize an LLM's ability to accurately respond to security-related inputs as stemming from security ''knowledge'', even though as probabilistic algorithms their behavior is likely quite different from human behavior in security contexts.}, meaning the ability to accurately respond to inputs in security contexts. Consequently, an active area of research is in techniques for measuring LLM security knowledge. Common measurement strategies include challenging LLMs with multiple choice security questions \cite{huang2025llms}, manually evaluating LLM responses to open-ended security questions \cite{chen2023can, prakash2025learned}, and designing benchmarks for measuring LLM performance on security tasks \cite{lee2026sec}. To target gap identification more precisely, corpora or benchmarks can be focused around specific areas or tasks. Researchers have also focused the search for knowledge gaps by reviewing user studies to identify the mental models or practices that manifest in LLM training data and result in security knowledge gaps \cite{chen2023can}.

All of these approaches require significant manual work and security expertise to design and implement. We seek a more efficient and broadly implementable method for identifying security areas in which LLM knowledge gaps exist. Such a method is complementary to other approaches; when it identifies an area of knowledge weakness, a more granular understanding of LLM shortcomings can be gained by applying existing assessment techniques. Indeed, such a method helps prioritize areas for additional measurement. In summary, we ask the following question: \\

\noindent
\textit{RQ: How can LLM knowledge gaps be efficiently identified by security area?} \\

\begin{figure*}
    \centering
    \includegraphics[width=\linewidth]{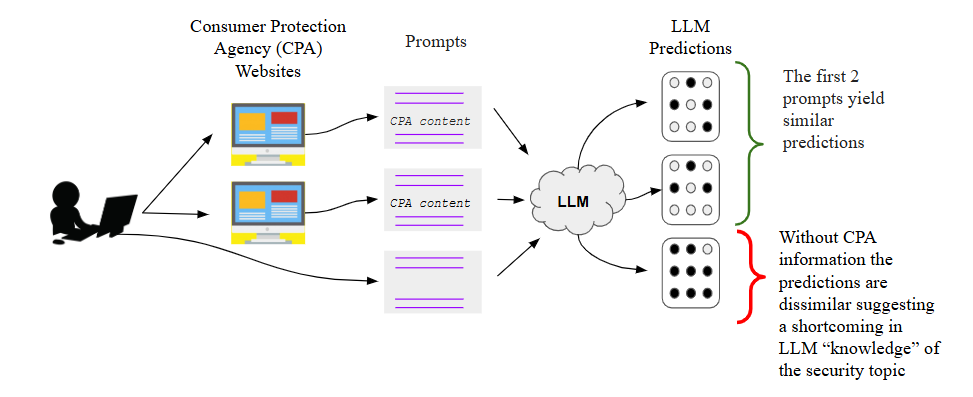}
    \caption{At a high level, the method of this paper looks for differences in LLM security performance when authoritative information is and is not present, to identify LLM knowledge gaps. The case studies instantiate security performance with the ability to identify security topics in consumer narratives, and authoritative information with content from Consumer Protection Agencies (CPAs), as shown in the figure. We find that if an LLM's predictions vary when authoritative information is or isn't present, this variation is indicative of poor performance in identifying the security topic (a ``knowledge gap'').}
    \label{fig:strategy}
\end{figure*}

The method introduced in this paper is inspired by techniques that use response instability to anticipate certain forms of LLM errors \cite{farquhar2024detecting}. Since LLM security errors may be due to pervasive misconceptions in training data, we do not rely upon instability in an LLM's response alone, but instead look for instability when prompts are augmented with authoritative information. In particular, our method uses security information from Consumer Protection Agencies (e.g., The Federal Trade Commission (FTC) or the Consumer Financial Protection Bureau (CFPB)) to test LLM response stability; if the LLM's response differs significantly with prompts that include CPA information from those without CPA information, this suggests the authoritative information is either not represented in, or differs from, the LLM's internal ``understanding'' of the topic, that is, the LLM has a knowledge gap. Figure~\ref{fig:strategy} provides a high-level illustration of the method.

We evaluate this method in the context of two prominent security problems, impostor scams and identity theft. Scams are a large and growing consumer safety risk, and impostor scams are reported by the FTC to be the most common form of scam \cite{ftc_top_scams, ftc-scam-report}. Impostor scams are characterized by the use of impersonation to enable scammers to benefit from pre-existing trust a target has in an institution (e.g., a bank) or a friend or family member, thus encouraging the target to take a financially harmful action. Identity theft is a financial harm in which a target's identity is appropriated to open new accounts in the person's name (e.g., credit card accounts) or gain access to existing accounts, without the target's permission.\footnote{While it is possible for an incident to involve both identity theft and an impostor scam, it is not required and CPAs treat them as separate risks.}

Impostor scams and identity theft are prominent consumer risks for which most Consumer Protection Agencies (CPAs) provide guidance. We use guidance from 6 CPAs to instantiate a prompt template instructing an LLM to evaluate a given text narrative for evidence of an impostor scam or identity theft. Using text narratives from the Consumer Financial Protection Bureau (CFPB) we built a data set of positive and negative examples of impostor scam narratives ($n=1357$) and a data set of positive and negative identity theft examples ($n=981$) to test the hypothesis that LLM response instability is indicative of poor LLM performance in identifying identity theft and impostor scams.

\begin{figure*}
    \centering
    \fbox{\includegraphics[width=.8\textwidth]{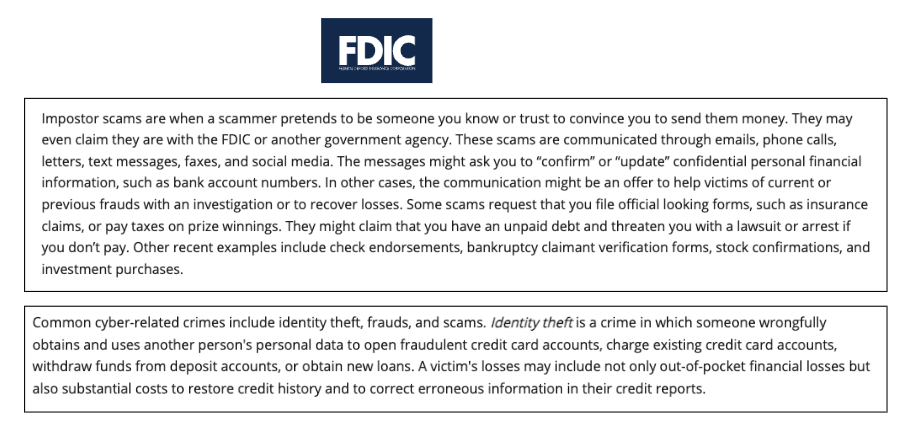}}
    \caption{The top and bottom rectangles include the FDIC text for the impostor scam and identity theft prompts (\textit{CPA-FDIC}), respectively. Both are found by querying the topic at \url{fdic.gov}, the impostor scam content is the second result and the identity theft content is the first result, when doing so as of the time of writing. All of the CPA content used in the experiments of this paper is published; the link is in Section~\ref{sec:cpa-prompts}.}
    \Description{FDIC text examples for impostor scam and identity theft prompts.}
    \label{fig:FDIC-two-examples}
\end{figure*}

Across 5 models in 2 leading LLM families (GPT\footnote{\url{https://developers.openai.com/api/docs/models}} and Gemini\footnote{\url{https://docs.cloud.google.com/vertex-ai/generative-ai/docs/models/}}) and the 2 consumer protection risks (impostor scams and identity theft) we demonstrate that the method is successful: if LLM predictions are stable across information from the 6 CPAs, then the LLM is able to accurately identify the security topic in CFPB complaints, but if LLM predictions are not stable, then the LLM does not accurately identify the security topic.

In summary we make the following contributions:
\begin{itemize}
\item \textbf{Semi-automated Identification of LLM Security Knowledge Gaps:} A broadly implementable method for identifying LLM knowledge gaps by security area. The method uses publicly available authoritative information and does not require labeled data. 
\item \textbf{2 Human-labeled Security Data Sets}: Curated data sets of CFPB narratives containing positive and negative examples of impostor scams ($n=1357$) and identity theft ($n=981$).
\item \textbf{Knowledge Test Results:} Outcome of the test on the 2 Data Sets, showing that 2 models, GPT 5.5 and Gemini 3.1 Pro, do not show evidence of an impostor scam knowledge gap, but all models appear to have identity theft knowledge gaps.
\end{itemize}

\subsection{Related Work}
\label{sec:relwork}

The problem of LLM knowledge gaps has been extensively studied in the context of individual questions or prompts. For example,
\cite{feng2024don} uses multiple LLMs to collaboratively identify when a question encounters a knowledge gap and decide whether an LLM should abstain from answering the question. Other techniques include eliciting confidence scores from LLMs \cite{tian2023just} or relying on some other form of judgment or ``self-reflection'' by the LLM \cite{kadavath2022language}. We seek a solution that doesn't depend on the LLM's awareness of the knowledge gap because security knowledge gaps may be associated with incorrect information that dominates training data, making it difficult for an LLM, or even a group of LLMs, to recognize.

Existing strategies for identifying LLM security knowledge gaps include multiple choice questions \cite{siddiq2025assessing, huang2025llms} or code vulnerability snippets \cite{sajadi2025llms} with which to challenge an LLM, and benchmarks for testing LLM and LLM agent ability to complete security tasks \cite{lee2026sec}. Others have approached the problem from a training data perspective, and used security misconceptions identified in user studies and commonly asked end-user security questions to focus an inquiry into LLM knowledge \cite{chen2023can,prakash2025learned}. This paper tackles the complementary problem of efficiently identifying \textit{areas} of security knowledge gaps. When such areas are identified, the existing strategies may guide model fine-tuning or help isolate problematic use cases.

The method of this paper is inspired by statistical techniques to predict LLM errors by measuring response instability (e.g., \cite{farquhar2024detecting}) and by a paradigm introduced in \cite{palla2025policy} of using authoritative information for prompt-based content moderation. We measure LLM response instability across prompts that do and do not contain authoritative information about a security topic, and find that instability is associated with knowledge gaps.

Finally, similar to the LLM-as-judge paradigm \cite{zheng2023judging}, the method of this paper relies upon multiple LLM responses to the same data set to determine the output. However, the goal of the method is to identify knowledge gaps for a given LLM and so the multiple responses are with respect to the same LLM but with different prompts, each based on authoritative information. In addition, the method seeks to identify LLM response \textit{instability} rather than to identify a correct response as in the LLM-as-judge paradigm.

\section{Data and Methodology}
\label{sec:data_methodology}

The focus of this paper is binary (i.e., pass/fail) tests of LLM knowledge of a security topic, with a pass indicating the LLM is appropriate for use cases relying on the security topic. To determine whether a test is functioning correctly we need an independent metric for measuring knowledge. For the case studies of this paper, that metric is an LLM's ability to recognize a security topic in text narratives. 

More formally, if an LLM acting as a binary classifier for a security topic, correctly identifies (aka ``predicts'') the security topic in text narratives, where correctness is measured via precision, recall, and the derived measure, F1 score \cite{manning2008introduction}, we say the LLM has ``knowledge'' of the security topic.

\begin{figure*}
    \centering
    \includegraphics[width=.8\linewidth]{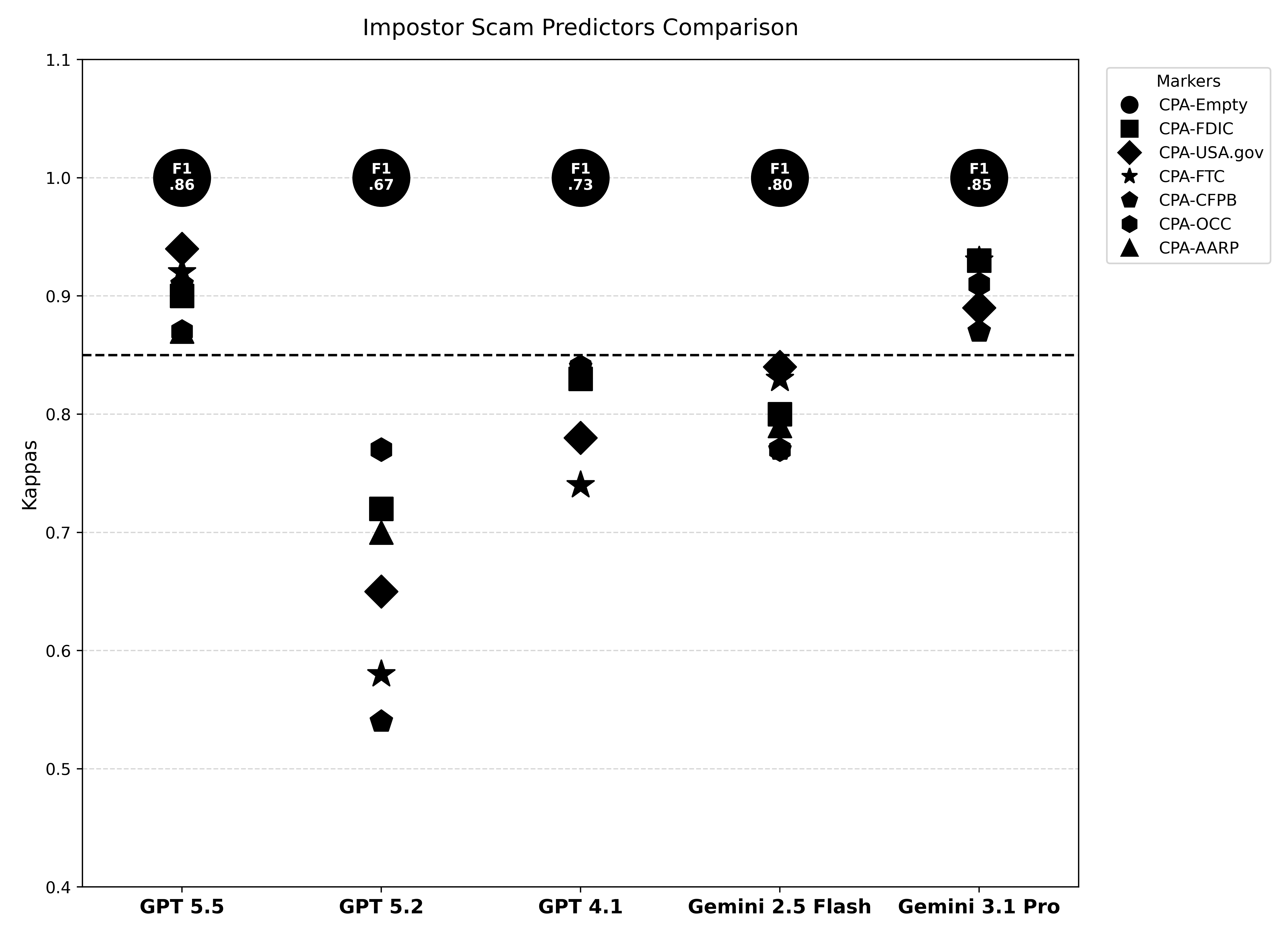}
    \caption{For the 5 LLMs, this diagram compares one run of each of the 6 CPA-based impostor scam prompts with individual runs of the baseline LLMs (the ``CPA-Empty'' impostor scam prompts). All CPA-based prompts produce predictions that are highly similar (Cohen's $\kappa$ greater than .85) to the predictions of GPT 5.5 and Gemini 3.1 Pro, with no additional information, and so both models pass the test. GPT 5.5 achieves precision of .82 and recall .9 and Gemini 3.1 Pro achieves precision of .79 and recall of .92. In contrast, CPA-based prompts all yield predictions that are less similar to their respective models ($.54\le{\kappa}\le{.84}$), and those LLMs have an average precision of $.61$. Each model's F1 score with the CPA-Empty impostor scam prompt is shown inside the circles at the top of the diagram ($\kappa=1$). The outcome of the second run of the same experiments was the same, only GPT 5.5 and Gemini 3.1 Pro passed.}
    \label{fig:scam-predictors}
\end{figure*}

We denote a data set of text narratives, as $\mathcal{D}$\,=\,$\{D_1, D_2,\ldots,D_n\}$, and the output of an LLM, $\mathcal{L}$, given prompt, $\mathcal{P}$, and data set, $\mathcal{D}$, as $\mathcal{L}(\mathcal{P}, \mathcal{D})$\,=\,$\{\mathcal{L}(D_1,\mathcal{P}),\dots,\mathcal{L}(D_n,\mathcal{P})\}$
where for all $D_i$ and $\mathcal{P}$, $\mathcal{L}(D_i,\mathcal{P})\in\{0,1\}$. We term $\mathcal{L}(\mathcal{P}, \mathcal{D})$ a \textit{predictor}. For simplicity of exposition, we often refer to the performance (e.g., precision and recall) of a predictor by referring to the prompt that defines it (i.e., the precision of $\mathcal{P}$ rather than the precision of $\mathcal{L}(\mathcal{P}, \mathcal{D})$).

The following subsections describe the specific data sets and prompts in this paper.

\subsection{Data Sets}
\label{sec:datasets}

In Section~\ref{sec:caseimpostor} and Section~\ref{sec:identitycasestudy} we evaluate the test of this paper on 2 security topics, impostor scams and identity theft.
To test knowledge of those topics, we built data sets of text narratives from a publicly available source, complaints submitted to the Consumer Financial Protection Bureau (CFPB), a United States government agency responsible for consumer protection in the financial sector.
Since the CFPB was created in 2011 through the Dodd-Frank Act\footnote{12 U.S. Code § 5491 - Establishment of the Bureau of Consumer Financial Protection}, the bureau has collected and monitored complaints from consumers related to financial safety in a variety of contexts (e.g., credit reporting, mortgage lending, automobile financing, student loans, scams and fraud). Consumers submitting complaints to the CFPB have the option of submitting a text narrative in addition to selecting from prepopulated options in several fields. If they submit a narrative and consent to making it public, it is redacted by the CFPB to reduce reidentification risk following the CFPB's ``scrubbing standard'' \cite{cfpb_personal}. We built 2 data sets from those in the CFPB database that include scrubbed narratives as described below. Both data sets are available at a link in Appendix~\ref{app:openscience}.

\subsubsection{Impostor Scams Data Set}
\label{sec:impostordata}

While scams are a significant online safety risk, they are a small portion of the complaints received by the CFPB. To more efficiently build a set of positive examples (impostor scams) and negative examples (non-impostor scams or fraud) we used the ``issue'' selected by consumers who consented to including their (redacted) narratives in the CFPB database as a filter. The issues available to complainants that are most related to scams are ``fraud or scam'' and ``Problem with fraud alerts or security freezes''.

The definitional distinction between scams and non-scam fraud is that in scams, the user (or, complainant) takes self-harming actions \cite{modic2013scam}. In particular, while both types of complaints may involve transactions considered fraudulent, in the case of a scam a user is tricked into authorizing the transaction themselves, whereas in non-scam fraud a bad actor authorizes the transaction (e.g., using stolen credentials). That said, the terms ``fraud'' and ``scam'' are often used interchangeably, and so we started with the $34,015$ public narratives labeled with either the ``Fraud or scam'' or ``Problem with fraud alerts or security freezes'' issue that were available on April 18, 2025 in the CFPB Complaints Database to build a data set of impostor scams and non-impostor fraud or scams.

\begin{figure*}
    \centering
    \includegraphics[width=.8\linewidth]{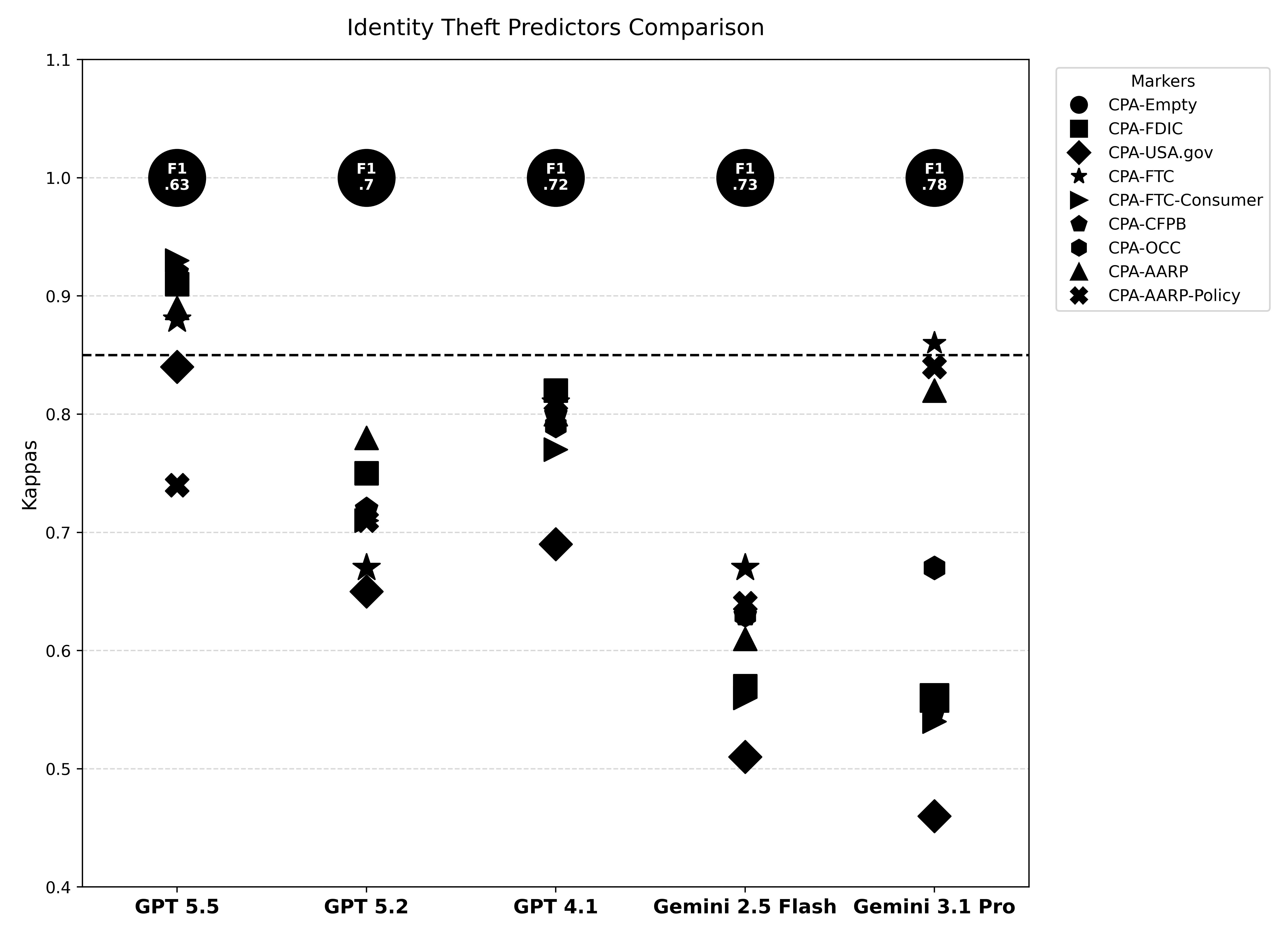}
    \caption{For the 5 LLMs, this diagram compares one run of each of the 6 CPA-based identity theft prompts with individual runs of the baseline LLMs (the ``CPA-Empty'' identity theft prompts). No model's baseline performance is very similar to all the CPA prompts, and 3 models are not very similar to any CPA prompts, hence none pass the test. The baseline models do not accurately detect identity theft in our test corpus; they achieve an average precision of $.606$. Each model's F1 score with the CPA-Empty identity theft prompt is shown inside the circles at the top of the diagram ($\kappa=1$). The results from the second runs of these experiments are quite similar with respect to placement above and below the $\kappa = .85$ line, although for GPT 5.5 the spread was greater; the bottom 2 kappa values are $.65$ and $.79$.}
    \label{fig:identity-theft-predictors}
\end{figure*}

Following the common qualitative research practice of using a codebook for data labeling \cite{macqueen1998codebook}, each of the 3 authors first independently labeled the same set of 100 narratives ``impostor scam'' or ``not impostor scam'', sampled from the $34,015$ public narratives, using an initial codebook consisting of a brief definition. The authors met to discuss cases of disagreement, clarify label definitions, and resolve disagreements. Based on the discussion, we updated and refined the codebook to expand on the definition and include positive and negative examples.
 
To assess whether all three authors had a shared understanding of the refined codebook, we independently labeled an additional shared set of 10 narratives and computed pairwise inter-rater agreement using Cohen's kappa \cite{cohen1960coefficient,gisev2013interrater}. The resulting pairwise kappas (0.80, 1.00, 0.80) indicated strong agreement across all authors and a consistent interpretation of the codebook definitions.
 
After achieving high inter-rater agreement, each author coded disjoint subsets of the data, resulting in a data set of $1357$ narratives after duplicate narratives were removed. During experiments in support of the test of Section~\ref{sec:test}, labeled samples were reviewed by all authors and $38$ labels were adjusted. The final labeled data set of $1357$ consists of $187$ impostor scams ($13.8\%$ of the data set). This data set is denoted as $\mathcal{D}_{IS}$ in the rest of this paper.

\subsubsection{Identity Theft Data Set}
\label{sec:identitydata}

While the term ``identity theft'' is often used broadly to refer to financial harms \cite{hoofnagle2007identity}, in this paper we follow the definition given by Congress in United States Code Title 18, Section 1028 \cite{uscode18-1028}; in short, the use of an individual's legal identification information to assume their identity for the purposes of creating or taking over new accounts and causing financial harm.

While identity theft continues to be a common financial harm \cite{ftc2024sentinel}, it is, like impostor scams, a relatively small portion of the CFPB Complaints database so we used a sampling strategy to identify positive examples of identity theft. The densest proportion of identity theft narratives appear to be in the 1,722 CFPB complaints for which complainants selected the label ``Identity theft / Fraud / Embezzlement'', however, that label was deprecated in late 2017 and so we augmented an initial sample from that label with narratives from product areas such as credit cards, which are often involved in identity theft. Our starting data pool had approximately 2000 narratives. As when building the impostor scam data set, each of the 3 authors first independently labeled the same set of 100 narratives ``identity theft'' or ``not identity theft'' using an initial codebook consisting of a brief definition. The authors met to discuss cases of disagreement, clarify label definitions, and resolve disagreements. Based on the discussion, we updated and refined the codebook to expand on the definition and include positive and negative examples.
 
To assess whether all three authors had a shared understanding of the refined codebook, we independently labeled an additional shared set of 10 narratives and achieved perfect agreement across all authors. Finally, each author coded disjoint subsets of the data, and duplicate narratives were removed, resulting in a data set of $981$ narratives, consisting of $427$ narratives describing identity theft ($43.5\%$ of the data set). This data set is denoted as $\mathcal{D}_{IT}$ in the rest of this paper.

\subsection{Prompts}
\label{sec:cpa-prompts}

The prompt engineering strategy of this paper incorporates authoritative safety guidance from Consumer Protection Agencies (CPAs) in the following simple template. We use this template for both the impostor scam and identity theft experiments. In the former, the second sentence of the prompt is ``Your task is to determine whether a consumer complaint describes an **Imposter Scam**'', and in the later the second sentence is ``Your task is to determine whether a consumer complaint describes **Identity Theft**''.

\begin{lstlisting}[language=Python]
LLM_instruction = ("You are a professional fraud detection analyst. Your task is to determine whether a consumer complaint describes {an **Imposter Scam**, **Identity Theft**}."
 {CPA Definition}
 "Determine whether the following consumer complaint describes {an imposter scam, identity theft}. Answer only 'yes' or 'no' without explanations."
{Consumer complaint: {narrative}})
\end{lstlisting}

In this paper, we term any instantiation of this template that uses only publicly-available, CPA-authored information, a \textit{CPA prompt}. We refer to a specific instantiation of this template by the organization that authored the CPA definition, e.g. CPA-FTC. If the template is used with no CPA definition we refer to the prompt as CPA-Empty.

This template does not assume substantial prompt engineering expertise. Indeed the only aspects of prompt engineering best practices present in the template that recent studies have found to not be commonly used \cite{jin2025understanding} are a role (``professional fraud detection analyst'') and output formatting (``Answer only `yes' or `no'...'').

We use CPA content related to each security topic that is easily retrieved by querying the CPA's website for the safety topic. Figure~\ref{fig:FDIC-two-examples} shows the FDIC content used in prompts as an example.

We experimented with including the URL of a CPA definition along with the definition and did not find it was associated with significant performance difference. This is compatible with the fact that, while all the models we used do support retrieving URLs for context to augment the prompt, it is generally advised to explicitly direct the LLM to use the URL in the prompt (e.g., \cite{gemini-urlcontext}). The complete text of all the CPA content in our experiments is published at a link in Appendix~\ref{app:openscience}.

\section{Testing LLM Security Knowledge}
\label{sec:test}

The hypothesis behind the knowledge test of this paper is that if an LLM's predictions change when augmented with authoritative information about a security topic, then the LLM has knowledge gaps in the area of the authoritative information. The test is implemented with an LLM, a data set of narratives, a threshold indicating the required LLM stability to pass the test and CPAs from which to extract authoritative information. To measure prediction change, the method uses a measure of inter-rater agreement, as is often done when using LLMs for data annotation (e.g., \cite{aguda2024large}).

The following makes the testing procedure concrete.\\

\noindent
\textbf{Procedure for Testing LLM Security Knowledge}\\
\textit{Input:} An LLM, $\mathcal{L}$, a security topic, $T$, a threshold $\tau$, $0<\tau<1$, CPAs, $CPA_i,\ldots,CPA_{k}$, for some integer, $k$, and a data set of narratives, $\mathcal{D}$\\
\textit{Steps:}\\
\begin{enumerate}
    \item For $i=1,\ldots,k$, query $CPA_i$ for topic $T$ and extract security content, $S_i$ to instantiate the prompt template of Section~\ref{sec:cpa-prompts}, resulting in prompts $\mathcal{P}_1,\ldots,\mathcal{P}_k$.
    \item Generate LLM responses for each prompt and data set, $\mathcal{D}$ : $\mathcal{L}(\mathcal{P}_1,\mathcal{D}),\ldots,\mathcal{L}(\mathcal{P}_k,\mathcal{D})$
    \item $\forall$ $1\le{i}\le{k}$, calculate the inter-rater agreement, $\kappa$ (e.g., Cohen's $\kappa$~\cite{cohen1960coefficient}), $\kappa_i = \kappa(\mathcal{L}(\mathcal{P}_i,\mathcal{D}), \mathcal{L}(CPA-Empty,\mathcal{D}))$
\end{enumerate}
\textit{Output}: If for all $i=1,\ldots, k$, $\kappa_i\ge{\tau}$, then the knowledge test passes, otherwise it fails.\\

Note that the test does not require labeled narratives as the test looks for \textit{changes} in predictions and does not evaluate the accuracy of predictions. However, to test the hypothesis behind the test, we use the labeled data sets from Section~\ref{sec:datasets} to measure results for the impostor scam (Section~\ref{sec:caseimpostor}) and identity theft case studies (Section~\ref{sec:identitycasestudy}). For changes to be detected, the method does benefit from a data set that contains both positive and negative examples and for rare security events, this may require a large sample of narratives (more on this in Section~\ref{sec:conclusion}).

In the following we discuss the results of applying the procedure of this section for the impostor scam and identity theft topics. All prompt experiments with Gemini 2.5 Flash were conducted in November and December 2025, and experiments with GPT 4.1 were run in January 2026. Experiments with GPT 5.2 and 5.5 and with Gemini 3.1 Pro, were run in April and May of 2026. In keeping with the goal of developing an easily implemented method, experiments were run with the default parameters when possible.

\subsection{Case Study: Impostor Scams}
\label{sec:caseimpostor}

To evaluate the knowledge gap detection method of Section~\ref{sec:test} for the impostor scam topic we first identified impostor scam definitions from 6 CPAs: AARP\footnote{\url{aarp.org}. The AARP (formerly, the American Association of Retired Persons) is a nonprofit that provides consumer guidance and services.}, CFPB\footnote{\url{https://www.consumerfinance.gov}}, FDIC\footnote{\url{fdic.gov}}, FTC\footnote{\url{ftc.gov}}, Office of the Comptroller of the Currency (OCC)\footnote{\url{www.occ.gov}} and USA.gov\footnote{\url{usa.gov}. While USA.gov serves to aggregate government information and services it does include impostor scam and identity information that is distinct from the other CPAs in this study.}. Impostor scam definitions were identified by visiting each site and querying for the topic. The extracted definitions (published at a link in Appendix~\ref{app:openscience}) were used to populate 6 CPA prompts (Step 1 of the procedure in Section~\ref{sec:cpa-prompts}). As specified by Step 2, each prompt was input to the five LLMs resulting in output $\mathcal{L}_j(\mathcal{P}_1, \mathcal{D}_{IS}),\ldots,\mathcal{L}_j(\mathcal{P}_6, \mathcal{D}_{IS})$ for each of the five LLMs, $\mathcal{L}_1,\ldots,\mathcal{L}_5$ (GPT 5.5, 5.2, 4.1, and Gemini 2.5 Flash and 3.1 Pro).

Finally, as specified by step 3, for each LLM, Cohen's kappa was calculated to measure the similarity of CPA-based predictions to predictions made without any CPA information (CPA-Empty). That is, for a given model, $\mathcal{L}_j$, 6 kappa values, $\kappa_{i,j} = \kappa(\mathcal{L}_j(\mathcal{P}_i,\mathcal{D}_{IS}), \mathcal{L}(CPA-{Empty},\mathcal{D}_{IS}))$, $i=1,\ldots,6$ were calculated.

We made a conservative choice of $\tau =.85$ to reduce the risk of failing to identify areas of security gaps; if and only if $\kappa_{ij}\ge{.85}$, for all $i=1,\ldots,6$, the test passed for LLM $\mathcal{L}_j$. Cohen's kappa values above $.8$ are typically considered ``strong'' \cite{mchugh_interrater_2012}. 

A visualization of one run of the procedure for Impostor Scams is in Figure~\ref{fig:scam-predictors}. Two LLMs pass the test, GPT 5.5 and Gemini 3.1 Pro. For the other LLMs all the $\kappa$ values are below $.85$, and some are below .6, indicating an agreement with the baseline model performance (i.e., with the CPA-Empty prompt) that is considered ``weak'' \cite{mchugh_interrater_2012}. For 2 of the models that appear to have impostor scam knowledge gaps, GPT 5.2 and GPT 4.1, the addition of CPA information consistently improved prediction precision. For Gemini 2.5 Flash, the impact of CPA information is erratic; precision declines with information from the AARP, FDIC and OCC, and improves with information from the CFPB, FTC and USA.gov.

The procedure of Section~\ref{sec:test} was run twice for each LLM and the pass/fail outcome was the same for each LLM. That is, GPT 5.5 and Gemini 3.1 Pro, ``pass'' each time, while the others fail. Indeed, the outcome is the same across both runs for a range of $\tau$ values, $.78\le{\tau}\le{.85}$. 

The precision and recall measurements for both runs of the CPA prompts are in Table~\ref{tab:impostor_scams_metrics} and the kappa results are in Table~\ref{table:impostor-scam-kappas}.

\begin{table*}[htbp]
\centering
\resizebox{\textwidth}{!}{%
\begin{tabular}{lcccccccccccccc}
\toprule
Model & \multicolumn{2}{c}{\textit{CPA-Empty}} & \multicolumn{2}{c}{\textit{CPA-AARP}} & \multicolumn{2}{c}{\textit{CPA-CFPB}} & \multicolumn{2}{c}{\textit{CPA-FDIC}} & \multicolumn{2}{c}{\textit{CPA-FTC}} & \multicolumn{2}{c}{\textit{CPA-OCC}} & \multicolumn{2}{c}{\textit{CPA-USA.gov}} \\
\cmidrule(lr){2-3} \cmidrule(lr){4-5} \cmidrule(lr){6-7} \cmidrule(lr){8-9} \cmidrule(lr){10-11} \cmidrule(lr){12-13} \cmidrule(lr){14-15}
 & Precision & Recall & Precision & Recall & Precision & Recall & Precision & Recall & Precision & Recall & Precision & Recall & Precision & Recall \\
\midrule
\multirow{2}{*}{GPT 5.5} & 0.82 & 0.90 & 0.75 & 0.97 & 0.81 & 0.94 & 0.79 & 0.96 & 0.81 & 0.94 & 0.78 & 0.95 & 0.85 & 0.92 \\
 & 0.82 & 0.91 & 0.79 & 0.94 & 0.94 & 0.86 & 0.79 & 0.95 & 0.82 & 0.92 & 0.74 & 0.96 & 0.85 & 0.91 \\
\midrule
\multirow{2}{*}{GPT 5.2}& 0.52 & 0.96 & 0.79 & 0.88 & 0.92 & 0.75 & 0.77 & 0.89 & 0.87 & 0.76 & 0.63 & 0.93 & 0.81 & 0.81 \\
 & 0.51 & 0.95 & 0.76 & 0.89 & 0.89 & 0.76 & 0.71 & 0.89 & 0.87 & 0.76 & 0.63 & 0.93 & 0.8 & 0.94 \\
\midrule
\multirow{2}{*}{GPT 4.1} & 0.60 & 0.94 & 0.62 & 0.93 & 0.67 & 0.88 & 0.72 & 0.90 & 0.79 & 0.82 & 0.62 & 0.93 & 0.75 & 0.85 \\
 & 0.55 & 0.96 & 0.57 & 0.95 & 0.65 & 0.87 & 0.73 & 0.90 & 0.75 & 0.82 & 0.65 & 0.90 & 0.78 & 0.81 \\
\midrule
\multirow{2}{*}{Gemini 2.5 Flash} & 0.71 & 0.91 & 0.62 & 0.96 & 0.75 & 0.84 & 0.68 & 0.93 & 0.77 & 0.94 & 0.59 & 0.98 & 0.76 & 0.89 \\
 & 0.71 & 0.96 & 0.57 & 0.90 & 0.66 & 0.76 & 0.62 & 0.84 & 0.69 & 0.81 & 0.50 & 0.90 & 0.76 & 0.91 \\
\midrule
\multirow{2}{*}{Gemini 3.1 Pro} & 0.79 & 0.92 & 0.77 & 0.91 & 0.88 & 0.86 & 0.81 & 0.89 & 0.84 & 0.89 & 0.76 & 0.94 & 0.85 & 0.84 \\
 & 0.77 & 0.94 & 0.81 & 0.93 & 0.88 & 0.86 & 0.80 & 0.88 & 0.84 & 0.88 & 0.74 & 0.94 & 0.85 & 0.84 \\
\bottomrule
\end{tabular}%
}
\caption{Precision and recall of the impostor scam experiments and all CPA prompts and LLMs.}
\label{tab:impostor_scams_metrics}
\end{table*}

\subsection{Case Study: Identity Theft}
\label{sec:identitycasestudy}

Again following the knowledge gap detection method of Section~\ref{sec:test} we first identified identity theft definitions from 6 CPAs: AARP, CFPB, FDIC, FTC, OCC and USA.gov. As mentioned in Section~\ref{sec:identitydata} the distinction between identity theft and fraud is a subtle, legal issue and perhaps because of this, we found that identity theft was covered in 2 places for the AARP (\url{aarp.org} and \url{policybook.aarp}) and the FTC (\url{ftc.gov} and \url{consumer.ftc.gov}). Hence, we added 2 additional CPA prompts (\textit{CPA-AARP-Policy} and \textit{CPA-FTC-Consumer}) to accommodate information on identity theft from all 4 sites, for a total of 8 CPA prompts.

As specified by Step 2, each prompt was input to the five LLMs resulting in output $\mathcal{L}_j(\mathcal{P}_1, \mathcal{D}_{IT}),\ldots,\mathcal{L}_j(\mathcal{P}_8, \mathcal{D}_{IT})$ for each of the five LLMs, $\mathcal{L}_1,\ldots,\mathcal{L}_5$ (GPT 5.5, 5.2, 4.1 and Gemini 2.5 Flash and 3.1 Pro).

Finally, as specified by step 3, for each LLM, Cohen's kappa was calculated to measure the similarity of CPA-based predictions to predictions made without any CPA information (CPA-Empty). That is, for a given model, $\mathcal{L}_j$, 8 kappa values were calculated, $\kappa_{i,j} = \kappa(\mathcal{L}_j(\mathcal{P}_i,\mathcal{D}_{IT}), \mathcal{L}(CPA-Empty,\mathcal{D}_{IT}))$, $i=1,\ldots,8$ .

Using $\tau =.85$, none of the LLMs passed the test; Figure~\ref{fig:identity-theft-predictors} shows the results. For only GPT 5.5 and Gemini 3.1 Pro were any CPA predictors sufficiently similar to the CPA-Empty predictor ($\kappa\ge{.85}$) and for both LLMs, other predictors were not sufficiently similar. The other LLMs (GPT 5.2 and 4.1, and Gemini 2.5 Flash) did not have any CPA predictors meeting the $\tau$ threshold of similarity with CPA-Empty. Indeed, the outcome (no passed models) is the same for $.75\le{\tau}\le{.85}$ across both runs.

The conclusion of the test that the LLMs \textit{all} have an identity theft knowledge gap is compatible with the precision measurements of Table~\ref{tab:identity_theft_metrics_expanded}; the precision of the models ranges from $.5$ to $.73$. The test outcome is also compatible with existing research; \cite{prakash2025learned} also found evidence of LLM knowledge gaps in the area of identity theft. In particular, Prakash et al, found that GPT-4, Gemini 1.5 and Llama-3 all incorrectly endorsed VPNs for identity theft protection.

\begin{table*}[htbp]
\centering
\resizebox{\textwidth}{!}{%
\begin{tabular}{lcccccccccccccccccc}
\toprule
Model & \multicolumn{2}{c}{\textit{CPA-Empty}} & \multicolumn{2}{c}{\textit{CPA-AARP}} & \multicolumn{2}{c}{\textit{CPA-AARP-Policy}} & \multicolumn{2}{c}{\textit{CPA-CFPB}} & \multicolumn{2}{c}{\textit{CPA-FDIC}} & \multicolumn{2}{c}{\textit{CPA-FTC}} & \multicolumn{2}{c}{\textit{CPA-FTC-Consumer}} & \multicolumn{2}{c}{\textit{CPA-OCC}} & \multicolumn{2}{c}{\textit{CPA-USA.gov}} \\
\cmidrule(lr){2-3} \cmidrule(lr){4-5} \cmidrule(lr){6-7} \cmidrule(lr){8-9} \cmidrule(lr){10-11} \cmidrule(lr){12-13} \cmidrule(lr){14-15} \cmidrule(lr){16-17} \cmidrule(lr){18-19}
 & Precision & Recall & Precision & Recall & Precision & Recall & Precision & Recall & Precision & Recall & Precision & Recall & Precision & Recall & Precision & Recall & Precision & Recall \\
\midrule
\multirow{2}{*}{GPT 5.5} & 0.50 & 0.86 & 0.51 & 0.84 & 0.55 & 0.84 & 0.5 & 0.87 & 0.50 & 0.85 & 0.52 & 0.86 & 0.49 & 0.87 & 0.50 & 0.84 & 0.50 & 0.93 \\
 & 0.50 & 0.85 & 0.51 & 0.84 & 0.53 & 0.84 & 0.5 & 0.86 & 0.44 & 0.75 & 0.44 & 0.73 & 0.44 & 0.77 & 0.44 & 0.74 & 0.44 & 0.82 \\
\midrule
\multirow{2}{*}{GPT 5.2} & 0.57 & 0.90 & 0.54 & 0.90 & 0.63 & 0.87 & 0.53 & 0.92 & 0.53 & 0.92 & 0.52 & 0.94 & 0.52 & 0.93 & 0.62 & 0.86 & 0.52 & 0.95 \\
 & 0.57 & 0.89 & 0.53 & 0.91 & 0.63 & 0.85 & 0.54 & 0.93 & 0.53 & 0.91 & 0.51 & 0.93 & 0.52 & 0.93 & 0.61 & 0.88 & 0.52 & 0.96 \\
\midrule
\multirow{2}{*}{GPT 4.1} & 0.61 & 0.88 & 0.57 & 0.85 & 0.61 & 0.85 & 0.62 & 0.89 & 0.64 & 0.84 & 0.61 & 0.89 & 0.55 & 0.86 & 0.65 & 0.85 & 0.54 & 0.94 \\
 & 0.62 & 0.87 & 0.58 & 0.87 & 0.6 & 0.85 & 0.63 & 0.9 & 0.63 & 0.85 & 0.61 & 0.89 & 0.56 & 0.87 & 0.66 & 0.85 & 0.55 & 0.94 \\
\midrule
\multirow{2}{*}{Gemini 2.5 Flash} & 0.60 & 0.91 & 0.53 & 0.91 & 0.57 & 0.88 & 0.54 & 0.93 & 0.51 & 0.88 & 0.56 & 0.88 & 0.51 & 0.91 & 0.54 & 0.89 & 0.51 & 0.98 \\
 & 0.61 & 0.91 & 0.53 & 0.91 & 0.57 & 0.85 & 0.55 & 0.91 & 0.52 & 0.89 & 0.56 & 0.9 & 0.51 & 0.92 & 0.54 & 0.89 & 0.51 & 0.97 \\
\midrule
\multirow{2}{*}{Gemini 3.1 Pro} & 0.72 & 0.84 & 0.66 & 0.84 & 0.70 & 0.83 & 0.52 & 0.86 & 0.52 & 0.84 & 0.69 & 0.85 & 0.51 & 0.84 & 0.56 & 0.84 & 0.52 & 0.92 \\
 & 0.73 & 0.84 & 0.66 & 0.84 & 0.70 & 0.83 & 0.52 & 0.87 & 0.52 & 0.83 & 0.69 & 0.86 & 0.51 & 0.85 & 0.57 & 0.83 & 0.52 & 0.92 \\
\bottomrule
\end{tabular}%
}
\caption{Precision and recall of the identity theft experiments and all CPA prompts and LLMs.}
\label{tab:identity_theft_metrics_expanded}
\end{table*}

\begin{table*}
\centering
\begin{tabular}{ccccccc}
\textbf{Model} & \textit{\textbf{CPA-AARP}} & \textit{\textbf{CPA-CFPB}} & \textit{\textbf{CPA-FDIC}} & \textit{\textbf{CPA-FTC}} & \textit{\textbf{CPA-OCC}} & \textit{\textbf{CPA-USA.gov}}\\
\hline
GPT 5.5 & $.87$ & $.91$ & $.90$ & $.92$ & $.87$ & $.94$ \\
 & $.90$ & $.92$ & $.89$ & $.94$ & $.90$ & $.93$ \\
\hline
GPT 5.2 & $.70$ & $.54$ & $.72$ & $.58$ & $.77$ & $.65$ \\
 & $.67$ & $.55$ & $.68$ & $.55$ & $.76$ & $.62$ \\
\hline
GPT 4.1 & $.84$ & $.84$ & $.83$ & $.74$ & $.84$ & $.78$ \\
 & $.87$ & $.78$ & $.77$ & $.70$ & $.80$ & $.68$ \\
\hline
Gemini 2.5 Flash & $.79$ & $.77$ & $.80$ & $.83$ & $.77$ & $.84$ \\
 & $.74$ & $.70$ & $.70$ & $.70$ & $.66$ & $-.01$ \\
\hline
Gemini 3.1 Pro & $.93$ & $.87$ & $.93$ & $.93$ & $.91$ & $.89$ \\
 & $.92$ & $.86$ & $.90$ & $.89$ & $.91$ & $.85$ \\
\hline
\end{tabular}
\caption{Impostor Scam Experiments. Each numerical entry is the Cohen's Kappa, $\kappa$(CPA-Empty, *), where CPA-Empty is the baseline predictor for the model in that row and * is the CPA predictor for the respective column.}
\label{table:impostor-scam-kappas}
\end{table*}

\begin{table*}
\centering
\begin{tabular}{ccccccccc}
\textbf{Model} & \footnotesize{\textit{\textbf{CPA-AARP}}} & \footnotesize{\textit{\textbf{CPA-AARP-Policy}}} & \footnotesize{\textit{\textbf{CPA-CFPB}}} & \footnotesize{\textit{\textbf{CPA-FDIC}}} & \footnotesize{\textit{\textbf{CPA-FTC}}} & \footnotesize{\textit{\textbf{CPA-FTC-Consumer}}} & \footnotesize{\textit{\textbf{CPA-OCC}}} & \footnotesize{\textit{\textbf{CPA-USA.gov}}}\\
\hline
GPT 5.5 & $.89$ & $.74$ & $.92$ & $.91$ & $.88$ & $.93$ & $.91$ & $.84$\\
 & $.89$ & $.79$ & $.94$ & $.65$ & $.86$ & $.95$ & $.86$ & $.96$\\
\hline
GPT 5.2 & $.78$ & $.71$ & $.72$ & $.75$ & $.67$ & $.71$ & $.75$ & $.65$ \\
 & $.78$ & $.72$ & $.74$ & $.77$ & $.69$ & $.73$ & $.76$ & $-.01$ \\
\hline
GPT 4.1 & $.81$ & $.81$ & $.8$ & $.82$ & $.81$ & $.78$ & $.79$ & $.69$ \\
 & $.84$ & $.82$ & $.81$ & $.8$ & $.83$ & $.77$ & $.79$ & $.69$ \\
\hline
Gemini 2.5 Flash & $.61$ & $.64$ & $.63$ & $.57$ & $.67$ & $.56$ & $.63$ & $.51$\\
 & $.60$ & $.64$ & $.61$ & $.56$ & $.66$ & $.56$ & $.61$ & $.5$\\
\hline
Gemini 3.1 Pro & $.82$ & $.84$ & $.55$ & $.56$ & $.86$ & $.54$ & $.67$ & $.46$ \\
 & $.80$ & $.87$ & $.52$ & $.54$ & $.85$ & $.53$ & $.66$ & $.46$ \\
\hline
\end{tabular}
\caption{Identity Theft Experiments. Each numerical entry is the Cohen's Kappa, $\kappa$(CPA-Empty, *), where CPA-Empty is the baseline predictor for the model in that row and * is the CPA predictor for the respective column.}
\label{table:identity-theft-kappas}
\end{table*}

\section{Failure Modes, Generalizability and Abuse}
\label{sec:discussion}

The test of this paper uses security information from CPAs to gauge LLM ``knowledge’’ of security. The method assumes CPA information is sufficiently accurate and clear to support security topic detection and that models will pay attention to the CPA information supplied in prompts. In this section, we discuss the failure modes that arise when these assumptions are not met as well as the extent to which the test generalizes within security and can be abused.

\subsection{Imperfect CPA information}
\label{sec:imperfectcpa}

If the test is reliable, then CPA security content that is similar in meaning and clarity will yield detectors that perform similarly. The case study experiments suggest the test is reliable, although the CPA content is more clear and similar in the impostor scam case study than in the identity theft case study. This variation is associated with differences in the test results as discussed in this section.

The CPA definitions of impostor scams describe the same defining characteristic of impostor scams: impostor scams involve the impersonation of a person or organization that is already trusted by the target. The CPA detectors have modest spread in kappa scores with most models. In particular, for GPT 5.5, GPT 4.1, Gemini 2.5 Flash and Gemini 3.1 Pro, the kappas of CPA predictors are all within .07 of each other. The kappas of CPA predictors with GPT 5.2 have a larger range, [.54, .77] (Figure~\ref{fig:scam-predictors}).

In contrast, the spread in kappa scores for the identity theft detectors is generally larger. Gemini 2.5 Flash, GPT 4.1 and GPT 5.2 have the least spread in kappas, at .13. GPT 5.5 kappas have a spread of .19 and Gemini 3.1 Pro has the largest spread, .4 (Figure~\ref{fig:identity-theft-predictors}). The identity theft CPA definitions, while all correct, are also less similar than those for impostor scams. In particular, commensurate with their protection mission, they emphasize the effects of identity theft and don't clearly characterize identity theft itself. For example, while most CPAs mention fraudulent credit card charges as a potential outcome of identity theft, some connect such charges to \textit{new} accounts in a target's name whereas others allow for charges against existing accounts (fraud that does not require identity theft). Further, only the AARP-policy definition makes a distinction between identity theft (stealing personal information) and identity fraud (use of personal information to commit fraud). Again, while none of the CPA content is incorrect, they differ from each other and do not provide an unambiguous characterization of identity theft.

Hence, while it is certainly possible that CPA content and a pretrained model share the same flawed understanding of a security topic resulting in predictors with high kappas, and consequently, an incorrectly passed knowledge test, we do not see that in the case studies of this paper. While recent research has not found authoritativeness to be associated with sycophancy \cite{wang2026truth, sharma2024towards} the risk of predictors being unduly influenced by poor CPA content is managed by not disclosing the sources of the information provided and by incorporating several authoritative CPAs. In addition, while the goal of this paper is a test that is easy to implement, a subject matter expert could be used in a limited way to evaluate the quality of CPA information.

\subsection{Confident but Incorrect LLMs}
\label{sec:confident}

LLM predictions can vary in confidence. Lower confidence may manifest as response instability \cite{arana2026models}. Every experiment in this paper was run twice, with similar results, suggesting confident predictors. If a predictor is highly confident, CPA information that contradicts a prediction may not result in a change in response. There is not strong evidence of this phenomenon in the case studies, despite the apparent confidence of the predictors.

In the impostor scam case study, the 3 LLMs for which the CPA-Empty predictor achieves substantially less than .8 precision are GPT 5.2, GPT 4.1 and Gemini 2.5 Flash. In the GPT 5.2 and GPT 4.1 experiments, all the CPA predictors improve upon the precision of CPA-Empty, that is, despite apparent confidence in incorrect predictions, the presence of CPA information results in improved predictions. The results for the optimized model, Gemini 2.5 Flash, are mixed; in half of the CPA experiments precision improves over CPA-Empty, however the kappa values are all below .85 and half are at most .74, indicating that while the CPA predictors do not perform well, they are making different errors from the pretrained model. All the precision measurements are in Table~\ref{tab:impostor_scams_metrics}.

In the identity theft case study, the only model for which some CPA predictors are similar enough to the CPA-Empty predictor to potentially demonstrate this phenomenon is GPT 5.5. However, as discussed in Section~\ref{sec:imperfectcpa}, this may also be due to the shortcomings of the CPA content for identity theft. Indeed, the only CPA prompt that makes a clear distinction between identity theft and identity fraud, CPA-AARP-Policy, improves over the precision of the CPA-Empty predictor (precision of .55 and .53, versus .5 for CPA-Empty).

While there is not evidence of over-confident models ignoring CPA information in our experiments, the risk of this occurring may be reduced by incorporating techniques for eliciting confidence and/or reasoning, to force the predictor to explain how predictions are compatible with CPA information that may be at odds with the pretrained model’s internal understanding of the topic \cite{tian2023just, wei2022chain}.

\subsection{Generalizability}
\label{sec:generalizability}

The goal of this test is to determine whether a pre-trained LLM has sufficient knowledge of a category of security incident to recognize user text narratives describing potential incidents. We believe the case studies demonstrate the method shows promise for the wide range of security incidents for which CPA guidance is available (e.g., \cite{ftc2026techa}), and within this context may extend to text authored by others as well (e.g., journalists or security incident response professionals). There are, of course, many other forms of data that are used to detect security incidents (e.g., network traffic) for which there is not CPA content to guide a model, and the instantiated test of this paper does not apply, but the test paradigm may still be useful with other sources of authoritative information.

\subsection{Test Abuse}
\label{sec:abuse}

When making a test public there is a risk that future inputs could be modified to game the outcome of the test; for example, with knowledge of how a spam filter works, spammers may modify their email messages to avoid spam classification and thus reach more inboxes \cite{rao2012economics}. Influencing the outcome of this test requires either modifying the performance of the pretrained LLMs or the content provided by the CPAs, both of which require hard to obtain access and/or influence.

\section{Conclusion and Open Problems}
\label{sec:conclusion}

We have introduced a partially-automated method for identifying security areas in which LLMs may have knowledge gaps. This method does not rely on labeled data and is relatively easy to implement as it requires only authoritative information, for example, publicly available information from Consumer Protection Agencies. We applied the test to 2 areas of security (impostor scams and identity theft) and 5 LLMs, finding that exactly 2 models (GPT 5.5 and Gemini 3.1 Pro) demonstrate knowledge of impostor scams and all 5 models fail to demonstrate knowledge of identity theft. While these are imbalanced results, they are supported by recent (largely manual) experiments finding model shortcomings in many security areas, including identity theft. That is, the imbalanced model results are compatible with the existing evidence that security knowledge is not a strength of many LLMs \cite{prakash2025learned,chen2023can,meiklejohn2026helpbench,kim2026security}.

To the best of our knowledge, this paper presents the first broadly implementable test of LLM security knowledge, and many open problems remain. We highlight three areas of open problems.

\textbf{Additional experimentation.} Perhaps the biggest need is experimentation with more security contexts (beyond scams and identity theft). The test of this paper can be adapted to broadly test security knowledge using new security benchmarks (e.g., \cite{meiklejohn2026helpbench}) as the inputs (rather than consumer complaints), and look for stability of answers to benchmark questions in the presence of authoritative information on the question topics. As the security topics broaden, the authoritative information may come from non-CPA sources such as regulatory documents (e.g., for data privacy topics) or forums or research papers. 

In addition, more experiments runs would increase confidence in reliability. In \cite{prakash2025learned}, free text model responses were found to be semantically consistent on a sample of prompts over 5 runs, however, over 10 runs, \cite{kim2026security} found some ``contradictory'' responses to security and privacy prompts for several models. More work is needed to explore whether variation in the binary responses used by the test of this paper can impact test reliability over multiple runs.

\textbf{Authoritative content experimentation.} CPA guidance aims to protect consumers and may not be ideally suited to detecting LLM knowledge gaps. For example, as noted in Section~\ref{sec:identitycasestudy}, the identity theft CPA information is oriented towards identity theft markers and does not clearly distinguish identity theft from other harms, a distinction that is less relevant to someone seeking to recover from financial harm. Additional experimentation is needed to understand the characteristics (content and phrasing) of authoritative information that are associated with effective knowledge tests. This experimentation could result in guidance for CPAs as to how to craft guidance that best serves consumers and LLMs.

\textbf{Test refinement.} Another important open area is exploring method refinements. We have chosen a fixed threshold ($\tau=.85$) for Cohen's kappa, but the test outcomes are the same for a range of threshold values ($[.75, 1]$ for identity theft and $[.78, .85]$ for impostor scams). That said, other techniques for measuring inter-rater agreement could be required to diminish the influence of any one measurement (all measures can exhibit some paradoxical behavior \cite{mchugh_interrater_2012}). In addition, the spread of similarities to the baseline model performance (CPA-Empty) appears to be a useful indicator of knowledge gaps (Section~\ref{sec:imperfectcpa}); a combination of kappa thresholds and spread criterion may be more effective. 

Guidance for selecting a sufficiently large data set is also needed. We do not assume labeled data are available, but the method does require that the data set, $\mathcal{D}$, have both positive and negative examples, so that the measures of similarity between predictors are more meaningful. Security events are often rare and so test implementers may need to gather large data samples to be confident they have both positive and negative examples. In some contexts, this is an easier task than in others. For example, the CFPB groups various forms of fraud and scams together (Section~\ref{sec:data_methodology}), hence providing a data set in which the rate of impostor scams is much higher than in the overall complaints population. In other security contexts, there may be publicly disclosed data regarding incidents or vulnerabilities that can help ensure the test is applied to a pool with positive and negative examples.

\textbf{Remediation.} Finally, this paper does not explore how to integrate the detection of areas of knowledge gaps with remediation like model fine-tuning. We noted that in some cases of knowledge gaps, CPA information improves precision. It is worth exploring whether multiple CPA-based predictors can be combined via voting or other ensemble methods to generate a ``north star'' of correct labels that can help model developers overcome security gaps. This north star could yield examples that can be incorporated in dynamic security benchmarks as well.

\begin{acks} 
This work was partially funded by an unrestricted gift from Google’s Academic Research Awards (GARA) program.
\end{acks}

\bibliographystyle{ACM-Reference-Format}
\bibliography{references, sample}

\appendix 

\section{Open Science}
\label{app:openscience}

We have anonymously published our impostor scam and identity theft data sets at the following URL: \url{https://docs.google.com/spreadsheets/d/e/2PACX-1vRqRy7c5gNEiD-i-tNtWl8_XuJeP_9BazPQqRa1F7lOBJHX6SA6iZ_2Rke8QLby-A/pubhtml}

The CPA definitions used in the experiments and the human-authored codebooks are anonymously published at: \url{https://docs.google.com/document/d/e/2PACX-1vSJvvNEoNTL0Mm4l1f-pP1yCiTKOE1M5GpCAEwU7Tp-QNLmJjyJBqDE1DI0GKtWay7rwBjh50weENOE/pub}

The Cohen's kappa \cite{mchugh_interrater_2012}, precision, recall and F1 calculations \cite{manning2008introduction}, were implemented both in spreadsheets and notebooks using the formulae in the citations.

\section{Ethical Considerations}

This work introduces a method for detecting security ``knowledge'' gaps in LLMs. We recommend that the method be applied conservatively, meaning the LLM should be determined to perform similarly to detectors based on authoritative information from several independent sources before it is considered suitable for a task requiring knowledge in a given security area. We test our methods with content from 6 different authoritative sources (Consumer Protection Agencies) resulting in 6 impostor scam detectors and 8 identity theft detectors, to emphasize this point.

An additional ethical consideration is the data we use, the complaints data from the Consumer Financial Protection Bureau (CFPB) complaints database \cite{cfpb}. The database is publicly available and released through the CFPB’s Consumer Complaint Database \cite{cfpb}. Complaints are published only if the consumer opts in to share their narrative publicly at the time of submission, and consumers may withdraw this consent at any time. In addition, before narratives are published, the CFPB applies a narrative scrubbing process to remove personal information that could directly identify an individual \cite{cfpb_personal}. This process includes automated checks and human review to try to ensure that personal identifiers are not present in the publicly released data. Consumers are informed about the consent process and the CFPB’s review procedures before they choose to publish their narrative.

\section{Use of Generative AI}
The topic of this paper is security knowledge tests for LLMs and we have experimented with LLMs to develop the proposed test.

The prose of this paper was human-authored. We have occasionally used generative AI (primarily, Gemini and Claude) as an assistant in the following ways:
\begin{itemize}
\item To generate or debug code snippets for analysis or figure generation
\item To generate latex for tables and bibtex entries for references
\item To critique earlier paper drafts for clarity.
\end{itemize}

\end{document}